\documentclass[twocolumn,conference]{IEEEtran}
\usepackage[T1]{fontenc}
\usepackage[latin9]{inputenc}
\usepackage{amsmath}
\usepackage{amssymb}
\usepackage{graphicx}
\PassOptionsToPackage{normalem}{ulem}
\usepackage{ulem}
\usepackage[unicode=true,
 bookmarks=true,bookmarksnumbered=true,bookmarksopen=true,bookmarksopenlevel=1,
 breaklinks=false,pdfborder={0 0 0},pdfborderstyle={},backref=false,colorlinks=false]
 {hyperref}
\hypersetup{pdftitle={Your Title},
 pdfauthor={Your Name},
 pdfpagelayout=OneColumn, pdfnewwindow=true, pdfstartview=XYZ, plainpages=false}

\makeatletter

\providecommand{\tabularnewline}{\\}

\usepackage[caption=false,font=footnotesize]{subfig}

\@ifundefined{showcaptionsetup}{}{%
 \PassOptionsToPackage{caption=false}{subfig}}
\usepackage{subfig}
\makeatother

\begin{document}

\title{Frequency-Selective Hybrid Beamforming Based on Implicit CSI for
Millimeter Wave Systems}

\author{\IEEEauthorblockN{Hsiao-Lan~Chiang, Wolfgang~Rave, Tobias Kadur, and Gerhard~Fettweis}\IEEEauthorblockA{Vodafone Chair for Mobile Communications, Technische Universität Dresden,
Germany\\
Email: \{hsiao-lan.chiang, rave, tobias.kadur, gerhard.fettweis\}@ifn.et.tu-dresden.de}}
\maketitle
\begin{abstract}
Hybrid beamforming is a promising concept to achieve high data rate
transmission at millimeter waves. To implement it in a transceiver,
many references optimally adapt to a high-dimensional multi-antenna
channel but more or less ignore the complexity of the channel estimation.
Realizing that received coupling coefficients of the channel and pairs
of possible analog beamforming vectors can be used for analog beam
selection, we further propose a low-complexity scheme that exploits
the coupling coefficients to implement hybrid beamforming. Essentially,
the coupling coefficients can be regarded as implicit channel state
information (CSI), and the estimates of these coupling coefficients
yield alternatives of effective channel matrices of much lower dimension.
After calculating the Frobenius norm of these effective channel matrices,
it turns out that the effective channel having the largest value of
the Frobenius norm provides the solution to hybrid beamforming problem.
\end{abstract}

\section{Introduction}

With the rapid increase of data rates in wireless communications,
bandwidth shortage is getting more critical. Therefore, there is a
growing interest in using millimeter wave (mmWave) for future wireless
communications taking advantage of the enormous amount of available
spectrum \cite{Rappaport2014}. When systems operate at mmWave frequency
bands,  a combination of analog beamforming \cite{Hajimiri2005,Liberti1999}
and digital beamforming \cite{VanTrees2004} can be one of the low-cost
solutions, which is commonly called hybrid beamforming \cite{Ayach2012}-\nocite{Alkhateeb2016b}\nocite{Sohrabi2016}\cite{Chiang2016_PIMRC}.
Unquestionably, it is intractable to deal with hybrid beamforming
at a transmitter and a receiver simultaneously because there are four
unknown matrices (two analog and two digital beamforming matrices).
To avoid dealing with all these issues at the same time, one can simplify
the problem by initially assuming that the channel matrix is known.
Then the problem of hybrid beamforming on both sides (i.e., finding
the precoder and combiner) can be solved by utilizing the singular
value decomposition (SVD) of the channel matrix \cite{Ayach2012}-\nocite{Alkhateeb2016b}\nocite{Sohrabi2016}\cite{Chiang2016_PIMRC}.
Unfortunately, the overhead of some preliminary work, such as  channel
estimation for large-scale antenna arrays \cite{Chiang2016_ISWCS,Venugopal2017},
makes the previously proposed solutions difficult to obtain. 

Our previous work in \cite{Chiang2017_ICASSP} explains that the
analog beam selection based on the power estimates is equivalent to
the beam selection by the orthogonal matching pursuit (OMP) algorithm
\cite{Cai2011} when the analog beamforming vectors are selected from
orthogonal codebooks. However, the result of the analog beam selection
technique based on the received power can be further improved because
the factor dominating the performance of hybrid beamforming is the
singular values of the effective channel $\mathbf{H}_{E}$ rather
than the received power. A simple cure for the problem is that one
can reserve a few more candidates of the analog beamforming vectors
associated with the large received power levels. Then, the subset
of these candidates yielding maximum throughput will provide the optimal
solution to the hybrid beamforming. Again it is evident that the computational
complexity exponentially increases as the size of the enlarged candidate
set. Consequently, we have a strong motivation to find a relationship
between the observations for the analog beam selection and a key parameter
of the hybrid beamforming gain, and then use the relationship to facilitate
the analog beam selection and the corresponding optimal digital beamforming.
In the low SNR regime, we find that the Frobenius norm of $\mathbf{H}_{E}$
is the key parameter of the hybrid beamforming gain. Moreover, the
observations for the analog beam selection can be used to generate
$\mathbf{H}_{E}$. Accordingly, the collected observations yielding
the maximum value of the key parameter gives us the necessary information
to optimally select the analog as well as the corresponding digital
beamforming matrices.
\begin{figure*}[t]
\begin{centering}
\includegraphics[scale=0.75]{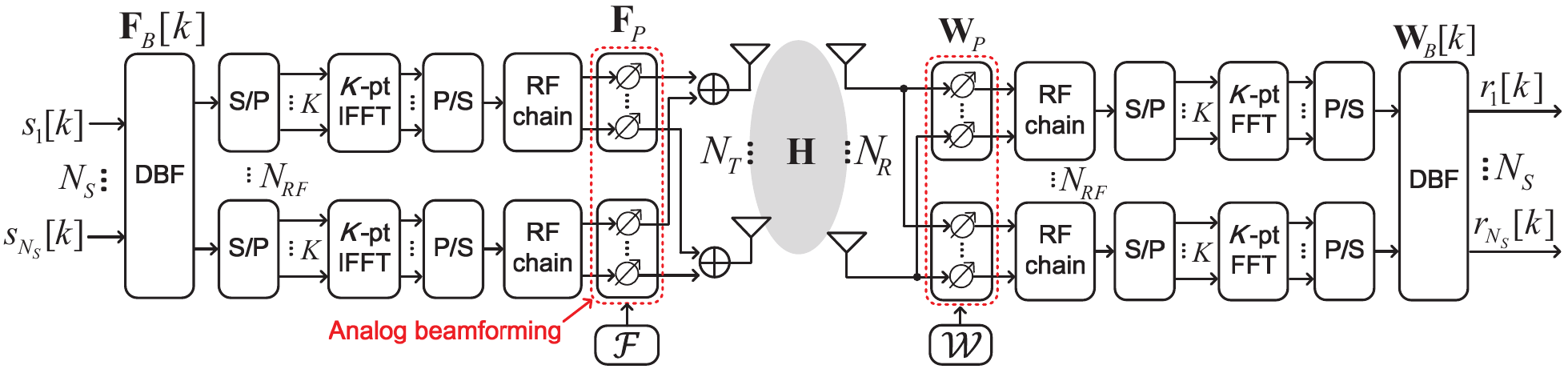}\caption{A transceiver has hybrid analog-digital beamforming structures on
both sides, where DBF is short for digital beamforming. Each analog
beamforming vector has multiple phase shifters connecting to one RF
chain, which includes a digital-to-analog converter (DAC) at the transmitter
or an analog-to-digital converter (ADC) at the receiver. \label{fig:System-diagram.} }
\par\end{centering}
{\small{}\vspace{-0.3cm}}{\small \par}
\end{figure*}

We use the following notations throughout this paper. $a$ is a scalar,
$\mathbf{a}$ is a column vector, and $\mathbf{A}$ is a matrix. \textbf{$\mathbf{a}_{n}$}
denotes the $n^{\text{th}}$ column vector of $\mathbf{A}$; $\left[\mathbf{A}\right]_{n,n}$
denotes the $n^{\text{th}}$ diagonal element of $\mathbf{A}$; $\left[\mathbf{A}\right]_{:,1:N}$
denotes the first $N$ column vectors of $\mathbf{A}$; $\left[\mathbf{A}\right]_{1:N,1:N}$
denotes the $N\times N$ submatrix extracted from the upper-left corner
of $\mathbf{A}$. $\mathbf{A}^{*}$, $\mathbf{A}^{H}$, and $\mathbf{A}^{T}$
denote the complex conjugate, Hermitian transpose and transpose of
$\mathbf{A}$. $\left\Vert \mathbf{A}\right\Vert _{F}$ denotes the
Frobenius norm of $\mathbf{A}$. $\mathbf{I}_{N}$ and $\mathbf{0}_{N\times M}$
denote the $N\times N$ identity and $N\times M$ zero matrices.

\section{System Model\label{sec:System_model}}

A system having a transmitter with an $N_{T}$-element uniform linear
antenna array (ULA) communicate $N_{S}$ OFDM data streams to a receiver
with an $N_{R}$-element ULA as shown in Fig. \ref{fig:System-diagram.}.
At the transmitter, a precoder $\mathbf{F}_{P}\mathbf{F}_{B}[k]$
at each subcarrier $k=1,\cdots,K$ includes a analog beamforming matrix
$\mathbf{F}_{P}\in\mathbb{C}^{N_{T}\times N_{RF}}$ and a digital
beamforming matrix $\mathbf{F}_{B}[k]\in\mathbb{C}^{N_{RF}\times N_{S}}$.
The $N_{RF}$ analog beamforming vectors in matrix $\mathbf{F}_{P}$
are selected from a predefined codebook $\mathcal{F}=\{\tilde{\mathbf{f}}_{n_{f}}\in\mathbb{C}^{N_{T}\times1},n_{f}=1,\cdots,N_{F}\}$
with the $n_{f}^{\text{th}}$ member represented as \cite{Liberti1999}
\begin{equation}
\tilde{\mathbf{f}}_{n_{f}}=\frac{1}{\sqrt{N_{T}}}\left[1,e^{j\frac{2\pi}{\lambda_{0}}\text{sin}(\phi_{T,n_{f}})\Delta_{d}},\cdots,e^{j\frac{2\pi}{\lambda_{0}}\text{sin}(\phi_{T,n_{f}})\cdot(N_{T}-1)\Delta_{d}}\right]^{T},\label{eq: f}
\end{equation}
where $\phi_{T,n_{f}}$ stands for the $n_{f}^{\text{th}}$ candidate
of the steering angles at the transmitter, $\Delta_{d}=\lambda_{0}/2$
is the distance between two neighboring antennas, and $\lambda_{0}$
is the wavelength at the carrier frequency. At the receiver, the combiner
$\mathbf{W}_{P}\mathbf{W}_{B}[k]$ has a similar structure as the
precoder: $\mathbf{W}_{P}\in\mathbb{C}^{N_{R}\times N_{RF}}$ and
$\mathbf{W}_{B}[k]\in\mathbb{C}^{N_{RF}\times N_{S}}$ are the analog
and digital beamforming matrices, respectively. Also, the columns
of $\mathbf{W}_{P}$ are selected from the other codebook $\mathcal{W}=\{\tilde{\mathbf{w}}_{n_{w}}\in\mathbb{C}^{N_{R}\times1},n_{w}=1,\cdots,N_{W}\}$,
where the members can be generated by the same rule as (\ref{eq: f}).
Due to hardware constraints, the analog beamforming matrices are constant
within one OFDM symbol duration. 

Via a coupling of the precoder, combiner, and a frequency-selective
fading channel $\mathbf{H}[k]\in\mathbb{C}^{N_{R}\times N_{T}}$,
the received signal $\mathbf{r}[k]\in\mathbb{C}^{N_{S}\times1}$ at
subcarrier $k$ can be written as 
\begin{equation}
\begin{alignedat}{1}\mathbf{r}[k] & =\sqrt{\rho}\cdot\mathbf{W}_{B}^{H}[k]\mathbf{W}_{P}^{H}\mathbf{H}[k]\underset{\mathbf{x}[k]}{\underbrace{\mathbf{F}_{P}\mathbf{F}_{B}[k]\mathbf{s}[k]}}+\underset{\mathbf{z}[k]}{\underbrace{\mathbf{W}_{B}^{H}[k]\mathbf{W}_{P}^{H}\mathbf{n}[k]}}\\
 & =\sqrt{\rho}\cdot\mathbf{W}_{B}^{H}[k]\mathbf{W}_{P}^{H}\mathbf{H}[k]\mathbf{x}[k]+\mathbf{z}[k],
\end{alignedat}
\label{eq: r}
\end{equation}
where $\rho$ stands for the average received power including the
transmit power, transmit antenna gain, receive antenna gain, and path
loss, $\mathbf{s}[k]\in\mathbb{C}^{N_{S}\times1}$ is the transmitted
pilot or data vector whose covariance matrix $\mathbf{R}_{s}=\text{E}[\mathbf{s}[k]\mathbf{s}^{H}[k]]$
is a diagonal matrix, and $\mathbf{n}[k]\in\mathbb{C}^{N_{R}\times1}$
is an $N_{R}$-dimensional independent and identically distributed
(i.i.d.) complex Gaussian random vector, i.e., $\mathbf{n}[k]\sim\mathcal{CN}(\boldsymbol{0}_{N_{R}\times1},\sigma_{n}^{2}\mathbf{I}_{N_{R}})$.
Furthermore, the precoded transmitted signal $\mathbf{x}[k]$ and
combined noise vector $\mathbf{z}[k]$ are enforced to satisfy the
following two conditions: constant transmit power at each subcarrier
and that the combined noise vector $\mathbf{z}[k]$ remains i.i.d.,
\begin{equation}
\begin{cases}
\text{tr}(\mathbf{R}_{x})=\text{tr}\left(\mathbf{F}_{P}\mathbf{F}_{B}[k]\mathbf{R}_{s}\mathbf{F}_{B}^{H}[k]\mathbf{F}_{P}^{H}\right)=\text{tr}(\mathbf{R}_{s}),\\
\mathbf{R}_{z}=\sigma_{n}^{2}\mathbf{W}_{B}^{H}[k]\mathbf{W}_{P}^{H}\mathbf{W}_{P}\mathbf{W}_{B}[k]=\sigma_{n}^{2}\mathbf{I}_{N_{S}},
\end{cases}\label{eq: constraint}
\end{equation}
which can also be regarded as power constraints on the precoder and
combiner.

The properties of mmWave channels have been widely studied recently
\cite{Rappaport2015,3GPP38900}. Based on the references, a simplified
cluster-based frequency-selective fading channel has $C$ clusters
and $R$ rays of each cluster. At subcarrier $k$, the channel matrix
can be written as
\begin{equation}
\begin{alignedat}{1}\mathbf{H}[k] & =\sum_{c=1}^{C}\sum_{r=1}^{R}\alpha_{c,r}\cdot e^{-j2\pi kl_{c,r}/K}\cdot\mathbf{a}_{A}(\phi_{A,c,r})\mathbf{a}_{D}(\phi_{D,c,r})^{H}\end{alignedat}
,
\end{equation}
where the channel characteristics are given by the following parameters:
$\alpha_{c,r}\in\mathbb{R}_{>0}$ describes the inter- and intra-cluster
power and $\sum_{c=1}^{C}\sum_{r=1}^{R}|\alpha_{c,r}|^{2}=1$. $l_{c,r}\in\mathbb{R}_{\geq0}$
stands for the delay index measured in unit of the sampling interval.
Intra-cluster angle of departure (AoD) $\begin{alignedat}{1}\phi_{D,c,r} & =\phi_{D,c}+c_{D}\Delta_{r}\end{alignedat}
$, where the mean $\phi_{D,c}\sim\mathcal{U}(-\tfrac{\pi}{2},\tfrac{\pi}{2})$,
and the other two factors (the angle spread $c_{D}$ and the offset
angle $\Delta_{r}$) are given in \cite{3GPP38900} Table 7.5-3 and
Table 7.5-6. In the same way, one can generate intra-cluster angle
of arrival (AoA) $\phi_{A,c,r}$. The departure array response vector
$\mathbf{a}_{D}(\phi_{D,c,r})$ has entries of equal magnitude and
is a function of AoD represented as
\begin{multline}
\mathbf{a}_{D}(\phi_{D,c,r})=\frac{1}{\sqrt{N_{T}}}\left[1,e^{j\tfrac{2\pi}{\lambda_{0}}\text{sin}(\phi_{D,c,r})\Delta_{d}},\cdots,\right.\\
\left.e^{j\tfrac{2\pi}{\lambda_{0}}\text{sin}(\phi_{D,c,r})(N_{T}-1)\Delta_{d}}\right]^{T},\label{eq: a_d}
\end{multline}
and the arrival array response vector $\mathbf{a}_{A}(\phi_{A,c,r})$
has a similar form as (\ref{eq: a_d}).

\section{Problem Statement \label{sec:Problem-Statement}}

The objective of the precoder $\mathbf{F}_{P}\mathbf{F}_{B}[k]\,\forall k$
and the associated combiner $\mathbf{W}_{P}\mathbf{W}_{B}[k]\,\forall k$
is to achieve the maximum throughput across all subcarriers subject
to the power constraints. That is, we seek matrices that solve 
\begin{equation}
\begin{gathered}\begin{gathered}{\displaystyle \underset{\mathbf{F}_{P},\mathbf{W}_{P},(\mathbf{F}_{B}[k],\mathbf{W}_{B}[k])\,\forall k}{\max}}\,\sum_{k=0}^{K-1}I(\mathbf{F}_{P},\mathbf{W}_{P},\mathbf{F}_{B}[k],\mathbf{W}_{B}[k]),\end{gathered}
\\
\text{s.t. }\begin{cases}
\mathbf{f}_{P,n_{rf}}\in\mathcal{F},\mathbf{w}_{P,n_{rf}}\in\mathcal{W}\:\:\forall n_{rf},\\
\text{tr}\left(\mathbf{F}_{P}\mathbf{F}_{B}[k]\mathbf{R}_{s}\mathbf{F}_{B}^{H}[k]\mathbf{F}_{P}^{H}\right)=\text{tr}(\mathbf{R}_{s})\,\forall k,\\
\mathbf{W}_{B}^{H}[k]\mathbf{W}_{P}^{H}\mathbf{W}_{P}\mathbf{W}_{B}[k]=\mathbf{I}_{N_{S}}\,\forall k,
\end{cases}
\end{gathered}
\label{eq: opt}
\end{equation}
where $\mathbf{f}_{P,n_{rf}}$ and $\mathbf{w}_{P,n_{rf}}$ are respectively
the $n_{rf}^{\text{th}}$ column vectors of $\mathbf{F}_{P}$ and
$\mathbf{W}_{P}$. The last two constraints are the consequences of
(\ref{eq: constraint}) and the throughput at subcarrier $k$ is defined
as \cite{Alkhateeb2016b}
\begin{multline}
I(\mathbf{F}_{P},\mathbf{W}_{P},\mathbf{F}_{B}[k],\mathbf{W}_{B}[k])=\\
\log_{2}\det\left(\mathbf{I}_{N_{S}}+\rho\mathbf{R}_{z}^{-1}\left(\mathbf{W}_{B}^{H}[k]\mathbf{W}_{P}^{H}\mathbf{H}[k]\mathbf{F}_{P}\mathbf{F}_{B}[k]\right)\right.\\
\left.\cdot\,\mathbf{R}_{s}\left(\mathbf{W}_{B}^{H}[k]\mathbf{W}_{P}^{H}\mathbf{H}[k]\mathbf{F}_{P}\mathbf{F}_{B}[k]\right)^{H}\right).\label{eq: I}
\end{multline}
The solution to (\ref{eq: opt}) is denoted as $(\mathbf{F}_{P,Opt},\mathbf{W}_{P,Opt},(\mathbf{F}_{B,Opt}[k],\mathbf{W}_{B,Opt}[k])\,\forall k)$. 

If explicit channel state information (CSI), $\mathbf{H}[k]$, $k=0,\cdots,K-1$,
is available, the problem of the precoder and combiner can be solved
according to the references \cite{Ayach2012,Alkhateeb2016b}. In this
paper, we consider a more pragmatic approach that channel knowledge
is neither given nor estimated. To efficiently get the solution to
(\ref{eq: opt}), we try an alternative expression of (\ref{eq: opt})
which has less probability to obtain the optimal solution as follows:
given two sets $\mathcal{I_{F}}$ and $\mathcal{I_{W}}$ containing
the candidates of $\mathbf{F}_{P}$ and $\mathbf{W}_{P}$, the achievable
data rate in (\ref{eq: opt}) is greater than or equal to

{\small{}\vspace{-0.3cm}
\begin{equation}
\begin{gathered}{\displaystyle \underset{\scriptsize\begin{array}{c}
\mathbf{F}_{P}\in\mathcal{I_{F}}\\
\mathbf{W}_{P}\in\mathcal{I_{W}}
\end{array}}{\max}}\underset{I_{LM}(\mathbf{F}_{P},\mathbf{W}_{P}):\text{ local maximum throughput}}{\underbrace{\left\{ \begin{gathered}\max_{(\mathbf{F}_{B}[k],\mathbf{W}_{B}[k])\,\forall k}\,\sum_{k=0}^{K-1}I(\mathbf{F}_{P},\mathbf{W}_{P},\mathbf{F}_{B}[k],\mathbf{W}_{B}[k])\\
\text{s.t. }\begin{cases}
\text{tr}\left(\mathbf{F}_{P}\mathbf{F}_{B}[k]\mathbf{R}_{s}\mathbf{F}_{B}^{H}[k]\mathbf{F}_{P}^{H}\right)=\text{tr}(\mathbf{R}_{s})\,\forall k\\
\mathbf{W}_{B}^{H}[k]\mathbf{W}_{P}^{H}\mathbf{W}_{P}\mathbf{W}_{B}[k]=\mathbf{I}_{N_{S}}\,\forall k
\end{cases}
\end{gathered}
\right\} }}.\end{gathered}
\label{eq: new}
\end{equation}
}These two versions will have the same data rate if $\mathcal{I_{F}}$
and $\mathcal{I_{W}}$ include $\mathbf{F}_{P,Opt}$ and $\mathbf{W}_{P,Opt}$
respectively. 

The reformulated problem in (\ref{eq: new}) becomes simpler because,
given $\mathbf{F}_{P}$ and $\mathbf{W}_{P}$, the \textit{inner}
problem (to obtain the local maximum throughput $I_{LM}(\mathbf{F}_{P},\mathbf{W}_{P})$)
is similar to conventional fully digital beamforming designs subject
to different power constraints \cite{VanTrees2004,Telatar1999}. In
simpler words, the critical issue of hybrid beamforming is to solve
the \textit{outer} problem by an additional maximization over all
members of $\mathcal{I_{F}}$ and $\mathcal{I_{W}}$. Therefore, the
motivation is to find $\mathcal{I_{F}}$ and $\mathcal{I_{W}}$, which
ideally include $\mathbf{F}_{P,Opt}$, $\mathbf{W}_{P,Opt}$, and
perhaps few other candidates, and then select a pair $(\mathbf{F}_{P},\mathbf{W}_{P})$
from $\mathcal{I_{F}}$ and $\mathcal{I_{W}}$ leading to the maximum
throughput.
\begin{figure*}[t]
\centering{}%
\begin{tabular}{ccc}
\subfloat[The achievable data rate by using the received power of the coupling
coefficients is 2.5 bit/s/Hz.]{\includegraphics[scale=0.3]{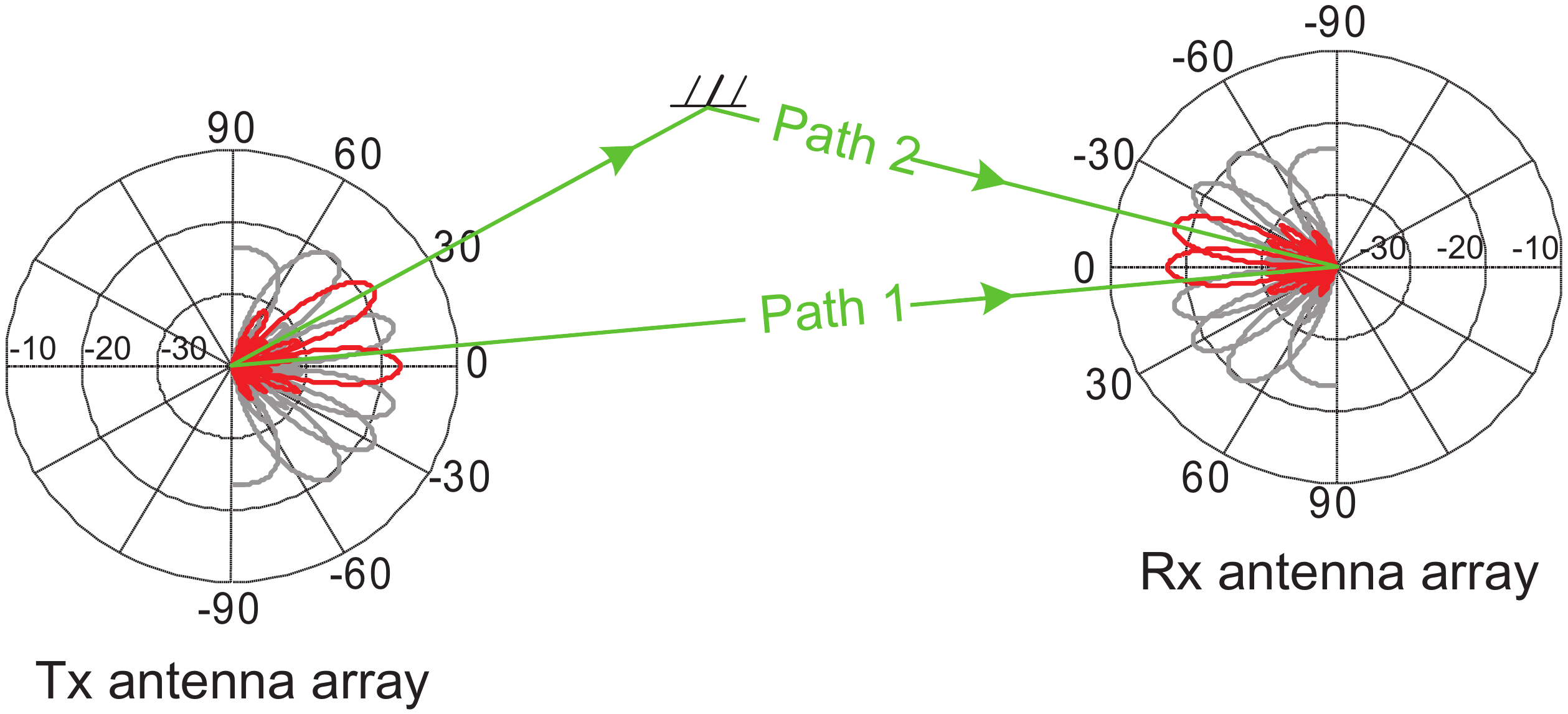}

} & $\qquad$  & \subfloat[The achievable data rate by using the linear combination of the two
analog beamforming vectors is 3 bit/s/Hz.]{\includegraphics[scale=0.3]{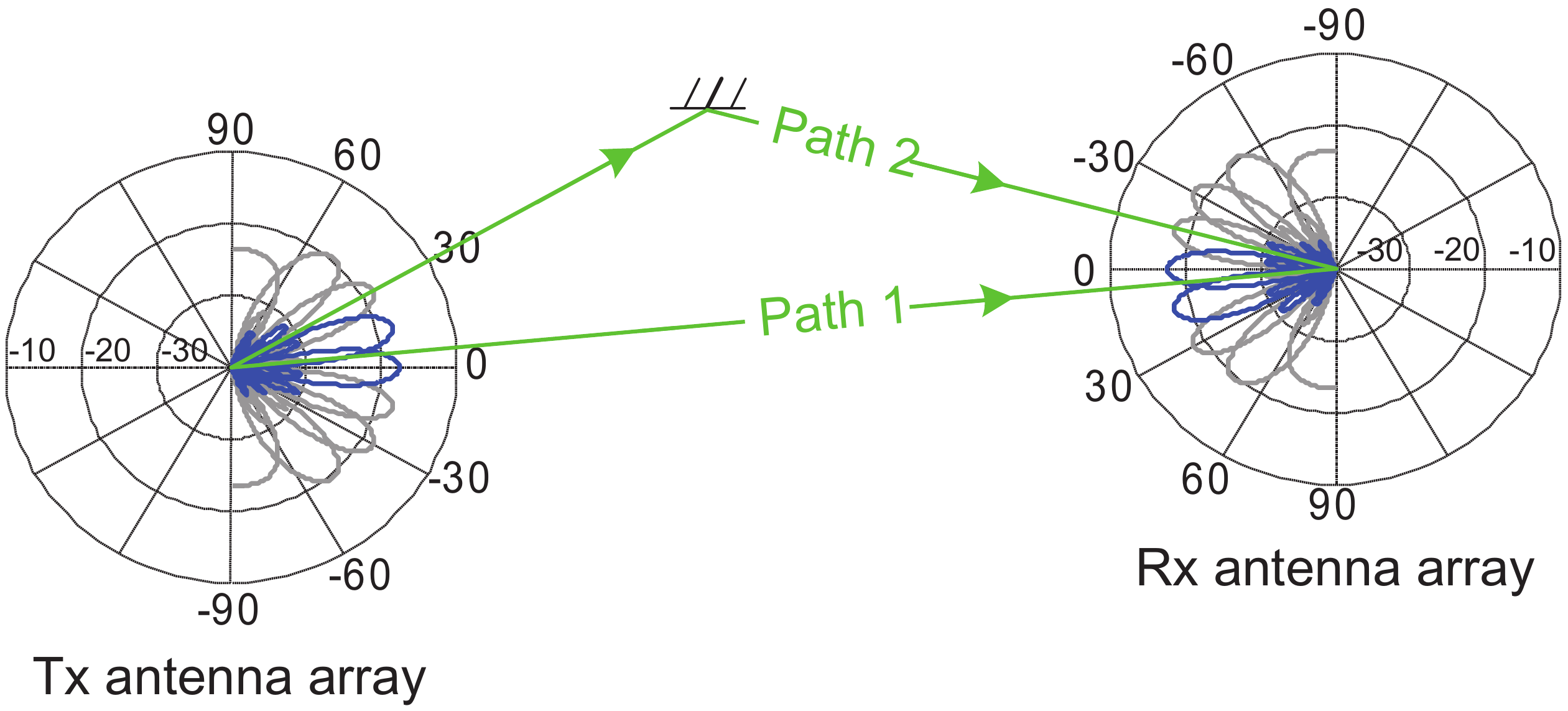}

}\tabularnewline
\end{tabular}\caption{A typical example of two different analog beam selection approaches.
In the simplified two-path channel model, the AoDs are $\{5^{\circ},30^{\circ}\}$,
the AoAs are $\{5^{\circ},-15^{\circ}\}$, and the difference in path
attenuation between path one and two amounts to 10 dB. \label{fig:example}}
\end{figure*}
\begin{figure}[t]
\centering{}\includegraphics[scale=0.45]{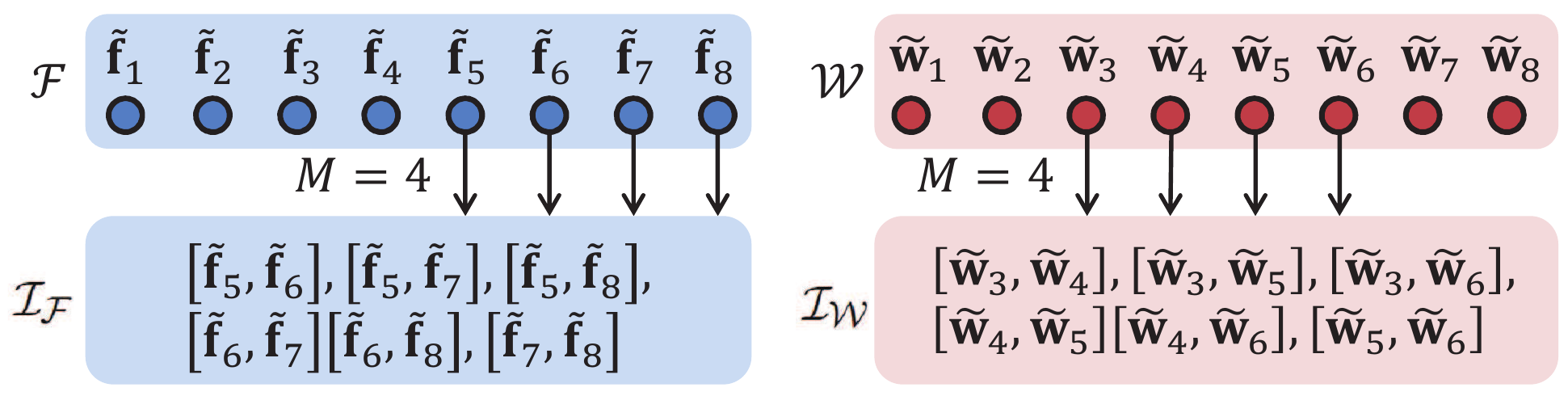}\caption{An example of the codebooks $\mathcal{F}$ and $\mathcal{W}$ with
eight columns and the sets $\mathcal{I_{F}}$ and $\mathcal{I_{W}}$
respectively consisting of $\binom{M}{N_{RF}}=\binom{4}{2}=6$ candidates
of $\mathbf{F}_{P}$ and $\mathbf{W}_{P}$.\label{fig:An-example-of-1} }
\end{figure}

\section{Hybrid Beamforming Algorithm Based on Implicit CSI\label{sec:HBF_algorithm}}

\subsection{Initial analog beam selection\label{subsec:Initial-analog-beam}}

To begin with, let us see how to obtain the sets $\mathcal{I_{F}}$
and $\mathcal{I_{W}}$ in (\ref{eq: new}) from the given codebooks
$\mathcal{F}$ and $\mathcal{W}$. We call this step \textit{initial
analog beam selection}. Since the hardware-constrained analog beamforming
matrices, $\mathbf{F}_{P}$ and $\mathbf{W}_{P}$, cannot be replaced
by the identity matrices, we have to train all or some of the columns
of the codebooks $\mathcal{F}=\{\tilde{\mathbf{f}}_{n_{f}},n_{f}=1,\cdots,N_{F}\}$
and $\mathcal{W}=\{\tilde{\mathbf{w}}_{n_{w}},n_{w}=1,\cdots,N_{W}\}$
by transmitting known pilot signals $\{s[k]\}_{k=0}^{K-1}$ satisfying
$|s[k]|^{2}=1\:\forall k$.

Then, an observation used for the analog beam selection at subcarrier
$k$ with respect to a trained beam pair $(\tilde{\mathbf{f}}_{n_{f}},\tilde{\mathbf{w}}_{n_{w}})$
is obtained by correlating the received signal with its transmitted
pilot
\begin{equation}
\begin{aligned}y_{n_{w},n_{f}}[k] & =\frac{s^{*}[k]}{|s[k]|^{2}}\underset{\text{received pilot}}{\underbrace{\left(\sqrt{\rho}\tilde{\mathbf{w}}_{n_{w}}^{H}\mathbf{H}[k]\tilde{\mathbf{f}}_{n_{f}}s[k]+\tilde{\mathbf{w}}_{n_{w}}^{H}\mathbf{n}[k]\right)}}\\
 & =\sqrt{\rho}\tilde{\mathbf{w}}_{n_{w}}^{H}\mathbf{H}[k]\tilde{\mathbf{f}}_{n_{f}}+z_{n_{w},n_{f}}[k],
\end{aligned}
\label{eq: y}
\end{equation}
where the effective noise $z_{n_{w},n_{f}}[k]\sim\mathcal{CN}(0,\sigma_{n}^{2})$
still has a Gaussian distribution with mean zero and variance $\sigma_{n}^{2}$
and similar observations become available on all subcarriers $k=0,\cdots,K-1$.
The observation $y_{n_{w},n_{f}}[k]$ can also be viewed as a coupling
coefficient of the channel and the beam pair $(\tilde{\mathbf{f}}_{n_{f}},\tilde{\mathbf{w}}_{n_{w}})$. 

Borrowing the idea from our previous works in \cite{Chiang2017_ICASSP,Chiang2017_WSA},
it shows that if the columns of $\mathcal{F}$ and $\mathcal{W}$
are orthogonal, the sum of the power of $K$ observations in one OFDM
symbol can be directly used for the analog beam selection. As a result,
$M$ (assume $M\geq N_{RF}$) analog beam pairs can be selected individually
and sequentially according to the sorted received power estimates 

\begin{equation}
\begin{gathered}(\hat{\mathbf{f}}_{P,m},\hat{\mathbf{w}}_{P,m})={\displaystyle \underset{\scriptsize\begin{array}{c}
\tilde{\mathbf{f}}_{n_{f}}\in\mathcal{F}\backslash\mathcal{F}',\tilde{\mathbf{w}}_{n_{w}}\in\mathcal{W}\backslash\mathcal{W}'\end{array}}{\arg\,\max}}\sum_{k=0}^{K-1}\left|y_{n_{w},n_{f}}[k]\right|^{2},\end{gathered}
\label{eq: ABF}
\end{equation}
where $m=1,\cdots,M$, $\mathcal{F}'=\{\hat{\mathbf{f}}_{P,n},n=1,\cdots,m-1\}$
and $\mathcal{W}'=\{\hat{\mathbf{w}}_{P,n},n=1,\cdots,m-1\}$ are
the sets consisting of the selected analog beamforming vectors from
iteration $1$ to $m-1$. We assume that $M\geq N_{RF}$ for the reason
that the first $N_{RF}$ selected analog beam pairs according to the
sorted magnitude of $\sum_{k=0}^{K-1}|y_{n_{w},n_{f}}[k]|^{2}$ may
not always lead to a good solution because we do not yet consider
the effect of digital beamforming during the analog beam selection
phase. To find the optimal solution, one has to further take into
account the linear combination of $N_{RF}$ analog beamforming vectors
selected from $\{\hat{\mathbf{f}}_{P,m}\,\forall m\}$ and $\{\hat{\mathbf{w}}_{P,m}\,\forall m\}$
with coefficients in digital beamforming. 

After selecting $M$ analog beam pairs, we define two sets $\mathcal{I_{F}}$
and $\mathcal{I_{W}}$ consisting of all the combinations of any $N_{RF}$
distinct analog beamforming vectors selected from $\{\hat{\mathbf{f}}_{P,m},m=1,\cdots,M\}$
and $\{\hat{\mathbf{w}}_{P,m},m=1,\cdots,M\}$, respectively, which
can be written as 
\begin{equation}
\begin{alignedat}{1}\mathcal{I_{F}} & =\{\overline{\mathbf{F}}_{P,i_{f}},i_{f}=1,\cdots,I_{F}\},\\
\mathcal{I_{W}} & =\{\overline{\mathbf{W}}_{P,i_{w}},i_{w}=1,\cdots,I_{W}\},
\end{alignedat}
\label{eq: beam space}
\end{equation}
where the cardinality $I_{F}=I_{W}=\binom{M}{N_{RF}}$ of both sets
is given by the binomial coefficient. The notation $\overline{\mathbf{F}}_{P,i_{f}}$
and $\overline{\mathbf{W}}_{P,i_{w}}$ respectively denote the $i_{f}^{\text{th}}$
and $i_{w}^{\text{th}}$ candidates of the analog beamforming matrices
$\mathbf{F}_{P}$ and $\mathbf{W}_{P}$. When $M$ becomes large,
there is a high probability that $\mathcal{I_{F}}$ and $\mathcal{I_{W}}$
include the global optimum solution $\mathbf{F}_{P,Opt}$ and $\mathbf{W}_{P,Opt}$. 

\textit{Schematic example}: let us consider a scenario with $N_{T}=N_{R}=8$
antenna elements, codebook sizes $N_{F}=N_{W}=8$, orthogonal codebooks
$\mathcal{F}=\mathcal{W}$ with steering angles given by $\{-90^{\circ}(\text{or }90^{\circ}),-48.59^{\circ},-30^{\circ},-14.48^{\circ},0^{\circ},14.48^{\circ},30^{\circ}$\\
$,48.59^{\circ}\}$, and $N_{RF}=2$ available RF chains to transmit
$N_{S}=2$ data streams at $\text{SNR}=5\text{ dB}$.

The channel realization as depicted in Fig. \ref{fig:example} has
two paths. First, in Fig. \ref{fig:example}(a), two analog beam pairs
highlighted in red are selected according to the sorted power level
estimates and steer towards these two paths. Before digital beamforming
comes into play, the analog beamforming vectors would be used with
the same weighting. If more than $N_{RF}=2$ analog beam pairs are
reserved according to the selection criterion expressed in (\ref{eq: ABF}),
more options with digital beamforming can be explored. In this example,
with $M=4$, we have $I_{F}=I_{W}=\binom{M}{N_{RF}}=\binom{4}{2}=6$
members in both $\mathcal{I_{F}}$ and $\mathcal{I_{W}}$, see Fig.
\ref{fig:An-example-of-1}. We can enumerate them explicitly as 
\[
\begin{alignedat}{1}\mathcal{I_{F}} & =\{\overline{\mathbf{F}}_{P,i_{f}},i_{f}=1,\cdots,6\},\\
\mathcal{I_{W}} & =\{\overline{\mathbf{W}}_{P,i_{w}},i_{w}=1,\cdots,6\}.
\end{alignedat}
\]
For instance, $\overline{\mathbf{F}}_{P,1}=[\tilde{\mathbf{f}}_{5},\tilde{\mathbf{f}}_{6}]$
and $\overline{\mathbf{W}}_{P,1}=[\tilde{\mathbf{w}}_{3},\tilde{\mathbf{w}}_{4}]$.
Then, one can try all 36 pairs of the members of $\mathcal{I_{F}}$
and $\mathcal{I_{W}}$ to determine the optimal weightings in digital
beamforming and choose the beam pairs yielding the maximum throughput,
which will be detailed in the following subsections. In general, there
will be a competition between spatial multiplexing gain over different
propagation paths and power gain available from the dominant path.
In this case, as shown in Fig. \ref{fig:example}(b) with the beams
highlighted in blue, the two analog beam pairs steering to the dominant
path lead to higher spectral efficiency. However, which beamforming
strategy has higher throughput in any specific case is not clear beforehand.

\subsection{Digital beamforming}

After the initial analog beam selection, we are in possession of the
two sets $\mathcal{I_{F}}$ and $\mathcal{I_{W}}$ that contain the
candidates of $\mathbf{F}_{P}$ and $\mathbf{W}_{P}$, and the objective
is to rapidly find which pair is the optimal. Before going into the
detail of our proposed scheme, let us review the relationship between
the analog and digital beamforming. Given one particular choice $(\overline{\mathbf{F}}_{P,i_{f}},\overline{\mathbf{W}}_{P,i_{w}})$
selected from the candidate sets $\mathcal{I_{F}}$ and $\mathcal{I_{W}}$,
it is clear that the goal of digital beamforming is to maximize the
local maximum throughput $I_{LM}(\overline{\mathbf{F}}_{P,i_{f}},\overline{\mathbf{W}}_{P,i_{w}})$
as defined in (\ref{eq: new}). As a consequence, for each subcarrier
the digital beamforming problem can be formulated as a throughput
maximization problem subject to the power constraints, which can be
stated as

\vspace{-0.3cm}

{\small{}
\begin{equation}
\begin{gathered}(\overline{\mathbf{F}}_{B,i}[k],\overline{\mathbf{W}}_{B,i}[k])={\displaystyle \underset{\mathbf{F}_{B}[k],\mathbf{W}_{B}[k]}{\arg\,\max}}I(\overline{\mathbf{F}}_{P,i_{f}},\overline{\mathbf{W}}_{P,i_{w}},\mathbf{F}_{B}[k],\mathbf{W}_{B}[k])\\
\text{s.t. }\begin{cases}
\text{tr}\left(\overline{\mathbf{F}}_{P,i_{f}}\mathbf{F}_{B}[k]\mathbf{R}_{s}\mathbf{F}_{B}^{H}[k]\overline{\mathbf{F}}_{P,i_{f}}^{H}\right)=\text{tr}(\mathbf{R}_{s}),\\
\mathbf{W}_{B}^{H}[k]\overline{\mathbf{W}}_{P,i_{w}}^{H}\overline{\mathbf{W}}_{P,i_{w}}\mathbf{W}_{B}[k]=\mathbf{I}_{N_{S}},
\end{cases}
\end{gathered}
\label{eq:constrained_DBF}
\end{equation}
}where $i=(i_{f}-1)I_{W}+i_{w}$ is the index specifying the combined
elements from $\mathcal{I_{F}}$ and $\mathcal{I_{W}}$.

In this problem, given ($\overline{\mathbf{F}}_{P,i_{f}},\overline{\mathbf{W}}_{P,i_{w}})$,
the corresponding optimal digital beamforming matrices are given by
(a detailed description of this part can be found in the journal version
of this paper \cite{Chiang2017_JSTSP})
\begin{equation}
\begin{alignedat}{1}\overline{\mathbf{F}}_{B,i}[k] & =\left(\overline{\mathbf{F}}_{P,i_{f}}^{H}\overline{\mathbf{F}}_{P,i_{f}}\right)^{-0.5}\left[\mathbf{V}_{E,i}[k]\right]_{:,1:N_{S}},\\
\overline{\mathbf{W}}_{B,i}[k] & =\left(\overline{\mathbf{W}}_{P,i_{w}}^{H}\overline{\mathbf{W}}_{P,i_{w}}\right)^{-0.5}\left[\mathbf{U}_{E,i}[k]\right]_{:,1:N_{S}},
\end{alignedat}
\end{equation}
where the columns of $\mathbf{V}_{E,i}[k]$ and $\mathbf{U}_{E,i}[k]$
are respectively the right and left singular vectors of the effective
channel

\vspace{-0.2cm}{\small{}
\begin{equation}
\begin{alignedat}{1}\mathbf{H}_{E,i}[k] & \;\triangleq(\overline{\mathbf{W}}_{P,i_{w}}^{H}\overline{\mathbf{W}}_{P,i_{w}})^{-0.5}\overline{\mathbf{W}}_{P,i_{w}}^{H}\mathbf{H}[k]\overline{\mathbf{F}}_{P,i_{f}}(\overline{\mathbf{F}}_{P,i_{f}}^{H}\overline{\mathbf{F}}_{P,i_{f}})^{-0.5}\\
 & \overset{\text{SVD}}{=}\mathbf{U}_{E,i}[k]\boldsymbol{\mathbf{\Sigma}}_{E,i}[k]\mathbf{V}_{E,i}^{H}[k].
\end{alignedat}
\label{eq:H_E}
\end{equation}
}{\small \par}

\subsection{Key parameter of hybrid beamforming gain\label{subsec:Key-parameter-of}}

Given a pair of elements selected from $\mathcal{I_{F}}$ and $\mathcal{I_{W}}$
and the corresponding optimal digital beamforming matrices across
all subcarriers, we can evaluate the local maximum throughput as
\begin{equation}
\begin{alignedat}{1} & I_{LM}(\overline{\mathbf{F}}_{P,i_{f}},\overline{\mathbf{W}}_{P,i_{w}})\\
 & =\sum_{k=0}^{K-1}\sum_{n_{s}=1}^{N_{S}}\log_{2}\left(1+\frac{\rho}{\sigma_{n}^{2}}\left[\boldsymbol{\mathbf{\Sigma}}_{E,i}^{2}[k]\right]_{n_{s},n_{s}}\left[\mathbf{R}_{s}\right]_{n_{s},n_{s}}\right),
\end{alignedat}
\label{eq: I_LM}
\end{equation}
where the diagonal elements of $\boldsymbol{\mathbf{\Sigma}}_{E,i}[k]$
are the singular values of the effective channel $\mathbf{H}_{E,i}[k]$.
Based on the candidate set $\{(\overline{\mathbf{F}}_{P,i_{f}},\overline{\mathbf{W}}_{P,i_{w}})\,|\,\forall i_{f},i_{w}\}$,
the pair leading to the maximum throughput provides the best approximation
of the global optimal analog beamforming matrices, denoted as $(\hat{\mathbf{F}}_{P},\hat{\mathbf{W}}_{P})$,
as well as the solution to the hybrid beamforming problem in (\ref{eq: new}),
\begin{equation}
\begin{alignedat}{1}\left(\hat{\mathbf{F}}_{P},\hat{\mathbf{W}}_{P}\right) & ={\displaystyle \underset{\scriptsize\begin{array}{c}
\overline{\mathbf{F}}_{P,i_{f}}\in\mathcal{I_{F}},\overline{\mathbf{W}}_{P,i_{w}}\in\mathcal{I_{W}}\end{array}}{\arg\,\max}}\,I_{LM}(\overline{\mathbf{F}}_{P,i_{f}},\overline{\mathbf{W}}_{P,i_{w}}).\end{alignedat}
\label{eq: max}
\end{equation}
However, this way of solving the problem requires the SVD of $\{\mathbf{H}_{E,i}[k]\}_{k=0}^{K-1}$
to obtain $I_{LM}(\overline{\mathbf{F}}_{P,i_{f}},\overline{\mathbf{W}}_{P,i_{w}})$
for each pair, which means that we have to repeat the calculation
as many as $\binom{M}{N_{RF}}^{2}$ times. 

To reduce the potentially large computational burden, we ask ourselves
what are the crucial parameter(s) or indicator(s) that actually determine
the throughput. To answer this question, let $\mathbf{R}_{s}=\frac{1}{N_{S}}\mathbf{I}_{N_{S}}$
(equal power allocation) so that the maximum achievable throughput
at subcarrier $k$  becomes
\begin{equation}
\begin{alignedat}{1} & I\left(\overline{\mathbf{F}}_{P,i_{f}},\overline{\mathbf{W}}_{P,i_{w}},\overline{\mathbf{F}}_{B,i}[k],\overline{\mathbf{W}}_{B,i}[k]\right)\\
 & \qquad\qquad=\sum_{n_{s}=1}^{N_{S}}\log_{2}\left(1+\gamma\left[\boldsymbol{\mathbf{\Sigma}}_{E,i}^{2}[k]\right]_{n_{s},n_{s}}\right),
\end{alignedat}
\label{eq: I_max}
\end{equation}
where $\gamma=\frac{\rho}{N_{S}\sigma_{n}^{2}}$ is the SNR. To find
the key parameter of the hybrid beamforming gain, we focus on the
low SNR regime. Using the fact that $\log(1+\gamma x)\approx\gamma x$
as $\gamma\rightarrow0$, the achievable data rate in (\ref{eq: I_max})
can be approximated by
\begin{equation}
\begin{alignedat}{1}I\left(\overline{\mathbf{F}}_{P,i_{f}},\overline{\mathbf{W}}_{P,i_{w}},\overline{\mathbf{F}}_{B,i}[k],\overline{\mathbf{W}}_{B,i}[k]\right) & \overset{\gamma\rightarrow0}{\approx}\gamma\sum_{n_{s}=1}^{N_{S}}\left[\boldsymbol{\mathbf{\Sigma}}_{E,i}^{2}[k]\right]_{n_{s},n_{s}}\\
 & \ \propto\sum_{n_{s}=1}^{N_{S}}\left[\boldsymbol{\mathbf{\Sigma}}_{E,i}^{2}[k]\right]_{n_{s},n_{s}}\\
 & \ \overset{}{\leq}\left\Vert \mathbf{H}_{E,i}[k]\right\Vert _{F}^{2}
\end{alignedat}
\label{eq: Frob_norm}
\end{equation}
with equality iff $N_{RF}=N_{S}$. For the case of $N_{RF}>N_{S}$,
$\left\Vert \mathbf{H}_{E,i}[k]\right\Vert _{F}^{2}$ corresponds
to the sum of all $N_{RF}$ (instead of only the $N_{S}$ strongest)
eigenvalues of $\mathbf{H}_{E,i}[k]\mathbf{H}_{E,i}^{H}[k]$. Assuming
that the sum of the weaker $N_{RF}-N_{S}$ eigenvalues of $\mathbf{H}_{E,i}[k]\mathbf{H}_{E,i}^{H}[k]$
is small, the approximation of $\sum_{n_{s}=1}^{N_{S}}\left[\boldsymbol{\mathbf{\Sigma}}_{E,i}^{2}[k]\right]_{n_{s},n_{s}}$
by $\left\Vert \mathbf{H}_{E,i}[k]\right\Vert _{F}^{2}$ seems to
be valid for most cases of interest. 

Fortunately, the matrix $\mathbf{H}_{E,i}[k]$ can be easily obtained
from the original observations $\{y_{n_{w},n_{f}}[k]\,\forall n_{w},n_{f},k\}$
\cite{Chiang2017_WSA}. Let us show the effective channel in (\ref{eq:H_E})
again and approximate the matrix $\overline{\mathbf{W}}_{P,i_{w}}^{H}\mathbf{H}[k]\overline{\mathbf{F}}_{P,i_{f}}$
by $\mathbf{Y}_{i}[k]$ (the elements of $\mathbf{Y}_{i}[k]$ can
be collected from the observations),

\vspace{-0.2cm}{\small{}
\begin{equation}
\begin{alignedat}{1}\mathbf{H}_{E,i}[k] & \triangleq(\overline{\mathbf{W}}_{P,i_{w}}^{H}\overline{\mathbf{W}}_{P,i_{w}})^{-0.5}\underset{\approx\mathbf{Y}_{i}[k]}{\underbrace{\overline{\mathbf{W}}_{P,i_{w}}^{H}\mathbf{H}[k]\overline{\mathbf{F}}_{P,i_{f}}}}(\overline{\mathbf{F}}_{P,i_{f}}^{H}\overline{\mathbf{F}}_{P,i_{f}})^{-0.5}\\
 & \approx(\overline{\mathbf{W}}_{P,i_{w}}^{H}\overline{\mathbf{W}}_{P,i_{w}})^{-0.5}\mathbf{Y}_{i}[k](\overline{\mathbf{F}}_{P,i_{f}}^{H}\overline{\mathbf{F}}_{P,i_{f}})^{-0.5}\\
 & \triangleq\hat{\mathbf{H}}_{E,i}[k].
\end{alignedat}
\label{eq: H_E_2}
\end{equation}
}For example, if $\overline{\mathbf{F}}_{P,i_{f}}=[\tilde{\mathbf{f}}_{P,1},\cdots,\tilde{\mathbf{f}}_{P,N_{RF}}]$
and $\overline{\mathbf{W}}_{P,i_{w}}=[\tilde{\mathbf{w}}_{P,1},\cdots,\tilde{\mathbf{w}}_{P,N_{RF}}]$,
where $\tilde{\mathbf{f}}_{P,n_{rf}}$ is the $n_{rf}^{\text{th}}$
column of $\mathcal{F}$ and $\tilde{\mathbf{w}}_{P,n_{rf}}$ is the
$n_{rf}^{\text{th}}$ column of $\mathcal{W}$, one has
\begin{equation}
\begin{alignedat}{1}\mathbf{Y}_{i}[k] & =\left[\begin{array}{ccc}
y_{1,1}[k] & \cdots & y_{1,N_{RF}}[k]\\
\vdots & \ddots & \vdots\\
y_{N_{RF},1}[k] & \cdots & y_{N_{RF},N_{RF}}[k]
\end{array}\right]\\
 & =\overline{\mathbf{W}}_{P,i_{w}}^{H}\mathbf{H}[k]\overline{\mathbf{F}}_{P,i_{f}}+\mathbf{Z}[k].
\end{alignedat}
\end{equation}
Therefore, given a pair $(\overline{\mathbf{F}}_{P,i_{f}},\overline{\mathbf{W}}_{P,i_{w}})$
selected from $\mathcal{I_{F}}$ and $\mathcal{I_{W}}$, we can rapidly
obtain the approximation of $\mathbf{H}_{E,i}[k]$, denoted as $\hat{\mathbf{H}}_{E,i}[k]$
in (\ref{eq: H_E_2}).

To conclude, the proposed solution can be stated as: first obtain
the candidate sets ($\mathcal{I_{F}}$ and $\mathcal{I_{W}}$) and
the approximation of $\mathbf{H}_{E,i}[k]$ ($\hat{\mathbf{H}}_{E,i}[k]$)
from the observations $\{y_{n_{w},n_{f}}[k]\,\forall n_{w},n_{f},k\}$
and then solve the maximization problem in (\ref{eq: max}), which
can be rewritten as
\begin{equation}
\begin{alignedat}{1}\left(\hat{i}_{f},\hat{i}_{w}\right) & ={\displaystyle \underset{\scriptsize\begin{array}{c}
\overline{\mathbf{F}}_{P,i_{f}}\in\mathcal{I_{F}},\overline{\mathbf{W}}_{P,i_{w}}\in\mathcal{I_{W}}\end{array}}{\arg\,\max}}\,I_{LM}(\overline{\mathbf{F}}_{P,i_{f}},\overline{\mathbf{W}}_{P,i_{w}})\\
 & \approx{\displaystyle \underset{\scriptsize\begin{array}{c}
\overline{\mathbf{F}}_{P,i_{f}}\in\mathcal{I_{F}},\overline{\mathbf{W}}_{P,i_{w}}\in\mathcal{I_{W}}\\
i=(i_{f}-1)I_{W}+i_{w}
\end{array}}{\arg\,\max}}\sum_{k=0}^{K-1}\left\Vert \hat{\mathbf{H}}_{E,i}[k]\right\Vert _{F}^{2}.
\end{alignedat}
\label{eq: ABS}
\end{equation}
According to the index pair $(\hat{i}_{f},\hat{i}_{w})$, we have
the selected analog and corresponding digital beamforming matrices
given by
\begin{equation}
\begin{alignedat}{1}\hat{\mathbf{F}}_{P} & =\overline{\mathbf{F}}_{P,\hat{i}_{f}},\\
\hat{\mathbf{W}}_{P} & =\overline{\mathbf{W}}_{P,\hat{i}_{w}},\\
\hat{\mathbf{F}}_{B}[k] & =(\hat{\mathbf{F}}_{P}^{H}\hat{\mathbf{F}}_{P})^{-0.5}\left[\hat{\mathbf{V}}_{E,\hat{i}}[k]\right]_{:,1:N_{S}},\\
\hat{\mathbf{W}}_{B}[k] & =(\hat{\mathbf{W}}_{P}^{H}\hat{\mathbf{W}}_{P})^{-0.5}\left[\hat{\mathbf{U}}_{E,\hat{i}}[k]\right]_{:,1:N_{S}},
\end{alignedat}
\label{eq: DBF-1}
\end{equation}
where $\hat{i}=(\hat{i}_{f}-1)I_{W}+\hat{i}_{w}$ and $\hat{\mathbf{U}}_{E,\hat{i}}[k]\hat{\boldsymbol{\mathbf{\Sigma}}}_{E,\hat{i}}[k]\hat{\mathbf{V}}_{E,\hat{i}}^{H}[k]=\text{SVD}(\hat{\mathbf{H}}_{E,\hat{i}}[k])$.
The complete proposed algorithm of the hybrid beamforming implementation
based on implicit CSI (i.e., the coupling coefficients $\{y_{n_{w},n_{f}}[k]\,\forall n_{w},n_{f},k\}$)
is shown in \textbf{Algorithm 1}, where $f(\hat{\mathbf{H}}_{E,i}[k])$
in Step 4 denotes the analog beam selection criterion by using (\ref{eq: I_max})
or (\ref{eq: Frob_norm}) with the argument $\hat{\mathbf{H}}_{E,i}[k]$
as

\vspace{-0.2cm}
\begin{flalign}
 & f\left(\hat{\mathbf{H}}_{E,i}[k]\right)\nonumber \\
 & =\begin{cases}
\sum_{n_{s}=1}^{N_{S}}\log_{2}\left(1+\gamma\left[\hat{\boldsymbol{\mathbf{\Sigma}}}_{E,i}^{2}[k]\right]_{n_{s},n_{s}}\right), & \text{w/o approx.}\\
\left\Vert \hat{\mathbf{H}}_{E,i}[k]\right\Vert _{F}^{2}, & \text{w/ approx.}
\end{cases}\label{eq: selection_criterion}
\end{flalign}
The advantages of the proposed algorithm are summarized as follows:
(1) we can omit high-dimensional channel estimation problems, and
(2) even though the cardinalities of $\mathcal{I_{F}}$ and $\mathcal{I_{W}}$
are large, the computational overhead is minor because we just need
to calculate the Frobenius norm of the effective channel matrices,
whose elements can be easily obtained from the observations $\{y_{n_{w},n_{f}}[k]\,\forall n_{w},n_{f},k\}$.

In (\ref{eq: I_max}), we simply assume that the transmit power is
equally allocated to $N_{S}$ data streams to facilitate the process
of finding the best value of the key parameter. Once the analog and
digital beamforming matrices are selected, the global maximum throughput
can be further improved by optimizing the power allocation (i.e.,
by a water-filling power allocation scheme \cite{Telatar1999}) for
$N_{S}$ data streams according to the effective channel condition
$\hat{\mathbf{H}}_{E,\hat{i}}[k]$ at each subcarrier.
\begin{figure}[t]
\hspace{-0.4cm}%
\begin{tabular}{|cl|}
\hline 
\multicolumn{2}{|l|}{\textbf{\small{}Algorithm 1: Hybrid beamforming based on implicit
CSI}}\tabularnewline
\hline 
\multicolumn{2}{|l|}{\textbf{\small{}Input: }{\small{}$\{y_{n_{w},n_{f}}[k]\,\forall n_{w},n_{f},k\}$}}\tabularnewline
\multicolumn{2}{|l|}{\textbf{\small{}Output: }{\small{}$\hat{\mathbf{F}}_{P}$, $\hat{\mathbf{W}}_{P}$,
$(\hat{\mathbf{F}}_{B}[k],\hat{\mathbf{W}}_{B}[k])\,\forall k$}}\tabularnewline
{\small{}1. } & {\small{}Given $\{y_{n_{w},n_{f}}[k]\,\forall n_{w},n_{f},k\}$, select
$M$ analog beam pairs}\tabularnewline
 & {\small{}$(\hat{\mathbf{f}}_{P,m},\hat{\mathbf{w}}_{P,m})$, where
$m=1,\cdots,M$, according to (\ref{eq: ABF}).}\tabularnewline
{\small{}2. } & {\small{}Generate two candidate sets $\mathcal{I_{F}}$ and $\mathcal{I_{W}}$
based on $\{\hat{\mathbf{f}}_{P,m}\,\forall m\}$ }\tabularnewline
 & {\small{}and $\{\hat{\mathbf{w}}_{P,m}\,\forall m\}$, respectively.}\tabularnewline
{\small{}3. } & {\small{}$\hat{\mathbf{H}}_{E,i}[k]=(\overline{\mathbf{W}}_{P,i_{w}}^{H}\overline{\mathbf{W}}_{P,i_{w}})^{-0.5}\mathbf{Y}_{i}[k](\overline{\mathbf{F}}_{P,i_{f}}^{H}\overline{\mathbf{F}}_{P,i_{f}})^{-0.5}$, }\tabularnewline
 & {\small{}where $\overline{\mathbf{F}}_{P,i_{f}}\in\mathcal{I_{F}}$,
$\overline{\mathbf{W}}_{P,i_{w}}\in\mathcal{I_{W}}$, and the entries
of $\mathbf{Y}_{i}[k]$}\tabularnewline
 & {\small{}are collected from $\{y_{n_{w},n_{f}}[k]\,\forall n_{w},n_{f}\}.$}\tabularnewline
{\small{}4. } & {\small{}$(\hat{i}_{f},\hat{i}_{w})={\displaystyle \underset{\scriptsize\begin{array}{c}
i=(i_{f}-1)I_{W}+i_{w}\end{array}}{\arg\,\max}\sum_{k=0}^{K-1}f(\hat{\mathbf{H}}_{E,i}[k]),}$}\tabularnewline
 & {\small{}where $f(\hat{\mathbf{H}}_{E,i}[k])$ is given in (\ref{eq: selection_criterion}). }\tabularnewline
{\small{}5. } & {\small{}\uline{Output}}{\small{}: $\hat{\mathbf{F}}_{P}=\overline{\mathbf{F}}_{P,\hat{i}_{f}}$
and $\hat{\mathbf{W}}_{P}=\overline{\mathbf{W}}_{P,\hat{i}_{w}}$.}\tabularnewline
{\small{}6. } & {\small{}$\hat{\mathbf{U}}_{E,\hat{i}}[k]\hat{\boldsymbol{\mathbf{\Sigma}}}_{E,\hat{i}}[k]\hat{\mathbf{V}}_{E,\hat{i}}^{H}[k]=\text{SVD}(\hat{\mathbf{H}}_{E,\hat{i}}[k])$,}\tabularnewline
 & {\small{}where $\hat{i}=(\hat{i}_{f}-1)I_{W}+\hat{i}_{w}$.}\tabularnewline
{\small{}7. } & {\small{}\uline{Output}}{\small{}: $\begin{cases}
\hat{\mathbf{F}}_{B}[k]=(\hat{\mathbf{F}}_{P}^{H}\hat{\mathbf{F}}_{P})^{-0.5}\left[\hat{\mathbf{V}}_{E,\hat{i}}[k]\right]_{:,1:N_{S}}\\
\hat{\mathbf{W}}_{B}[k]=(\hat{\mathbf{W}}_{P}^{H}\hat{\mathbf{W}}_{P})^{-0.5}\left[\hat{\mathbf{U}}_{E,\hat{i}}[k]\right]_{:,1:N_{S}}
\end{cases}$}\tabularnewline
\hline 
\end{tabular}
\end{figure}

\section{Simulation Results\label{sec:Simulation-Results}}

The system has $N_{T}=N_{R}=32$ antennas, $N_{RF}=2$ RF chains,
$N_{S}=2$ data streams, and $K=512$ subcarriers. The cluster-based
channel model has $C=5$ clusters including one line-of-sight (LoS)
and four NLoS clusters, and each cluster has $R=8$ rays. The codebooks
$\mathcal{F}$ and $\mathcal{W}$ have the same number of members,
$N_{F}=N_{W}=32$, and $32$ steering angle candidates are: $\left\{ \frac{180^{\circ}}{\pi}\cdot\sin^{-1}\left(\frac{\left(n_{f}-16\right)}{16}\right),\,n_{f}=1,\cdots,32\right\} $,
which yield orthogonal codebooks \cite{Chiang2016_WOWMOM}.

We chose the work in \cite{Ayach2012} that implements hybrid beamforming
based on the singular vectors of \textbf{$\mathbf{H}[k]\:\forall k$
}(i.e., \textit{explicit} \textit{CSI}) as a reference method for
comparison and extended it from single carrier to multiple carriers.
Different to the reference scheme, \textbf{Algorithm 1} uses the received
coupling coefficients (i.e., \textit{implicit} \textit{CSI}) as the
observations for the hybrid beamforming implementation. The received
coupling coefficients are commonly used for channel estimation \cite{Chiang2016_ISWCS,Mendez-Rial2016},
but in this paper we use them to directly implement the hybrid beamforming
on both sides. As a result, we can get rid of the overhead of channel
estimation.
\begin{figure}[t]
\centering{}\hspace*{-0.3cm}\includegraphics[scale=0.6]{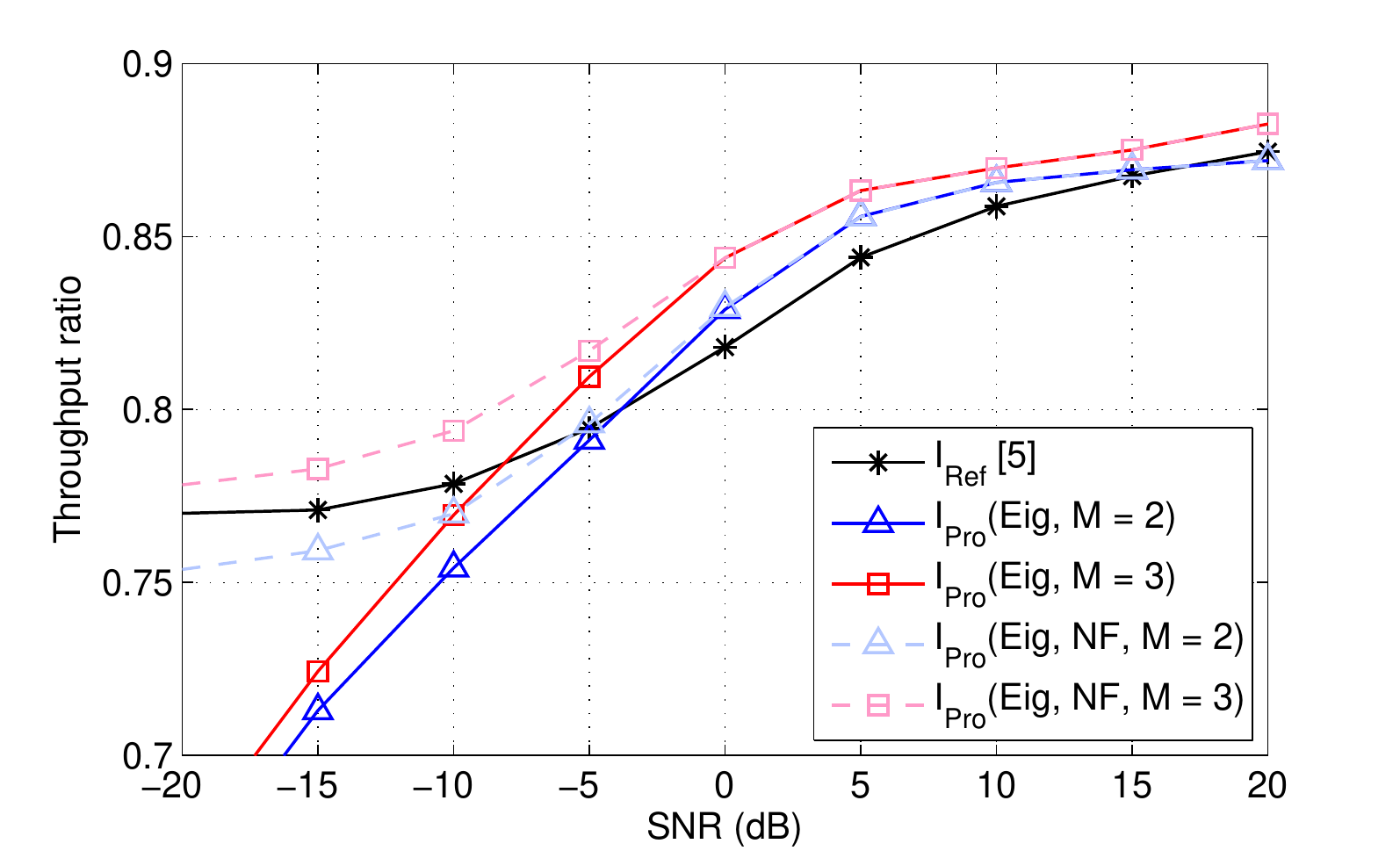}\caption{Normalized achievable throughput by the proposed and reference methods.
In the proposed method, different number $M$ of initially selected
analog beam pairs as well as different qualities of observations are
evaluated.\label{fig: orthogonal_ref_vs_eigen} }
\end{figure}
\begin{figure}[t]
\centering{}\hspace*{-0.3cm}\includegraphics[scale=0.6]{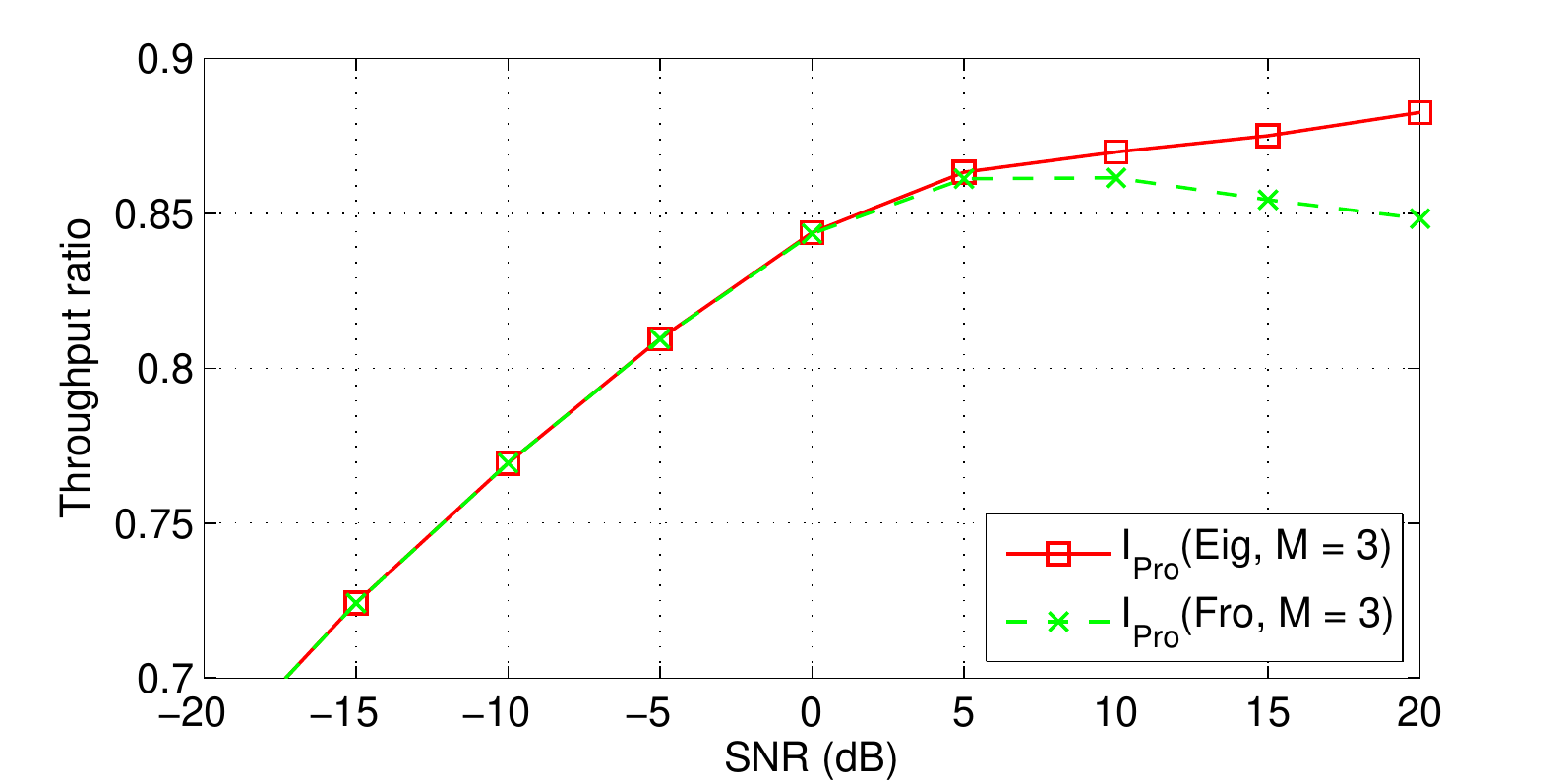}\caption{Curve $I_{Pro}(Eig,M=3)$ shown in Fig. \ref{fig: orthogonal_ref_vs_eigen}
and its approximation achieved by using the key parameter.\label{fig: orthogonal_eigen_vs_key} }
\end{figure}

To clearly present the difference in throughput by using different
methods, the calculated throughput values are normalized to the throughput
achieved by fully digital beamforming\footnote{The data rates achieved by fully digital beamforming used for the
normalization from $\text{SNR}=-20$ dB to $20$ dB (step by $5$
dB) are: $\{0.05,0.14,0.41,1.03,2.17,3.77,5.79,8.17,10.91\}$ in bit/s/Hz.}. Fig. \ref{fig: orthogonal_ref_vs_eigen} shows the achievable data
rates with $M=2,3$ initially selected analog beam pairs in the proposed
method (curves denoted as \textit{$I_{Pro}$}) and the reference approach
(curve \textit{$I_{Ref}$}). The data rates shown in \textit{$I_{Pro}(Eig,M=2,3)$}
are calculated by\textbf{
\begin{equation}
\frac{1}{K}\sum_{k=0}^{K-1}I(\hat{\mathbf{F}}_{P},\hat{\mathbf{W}}_{P},\hat{\mathbf{F}}_{B}[k],\hat{\mathbf{W}}_{B}[k]),\label{eq: DR}
\end{equation}
}where \textbf{$(\hat{\mathbf{F}}_{P},\hat{\mathbf{W}}_{P},\hat{\mathbf{F}}_{B}[k],\hat{\mathbf{W}}_{B}[k])$}
are the outputs of \textbf{Algorithm 1} with $f(\hat{\mathbf{H}}_{E,i}[k])=\sum_{n_{s}=1}^{N_{S}}\log_{2}(1+\gamma[\hat{\boldsymbol{\mathbf{\Sigma}}}_{E,i}^{2}[k]]_{n_{s},n_{s}})$
in Step 4. If the inputs of \textbf{Algorithm 1}, $\{y_{n_{w},n_{f}}[k]\,\forall n_{w},n_{f},k\}$,
do not take into account random noise signals, repeating the above-mentioned
steps leads to $I_{Pro}(Eig,N\!F,M\!=\!2,3)$\textit{.} $I_{Ref}$
is also calculated by (\ref{eq: DR}) with the solution of $(\mathbf{F}_{P},\mathbf{W}_{P},\mathbf{F}_{B}[k],\mathbf{W}_{B}[k])$
given in \cite{Ayach2012}. Furthermore, in Fig. \ref{fig: orthogonal_eigen_vs_key},
curve $I_{Pro}(Fro,M=3)$ uses the Frobenius norm of $\hat{\mathbf{H}}_{E,i}[k]$
(i.e., the key parameter of the hybrid beamforming gain) as a selection
criterion in \textbf{Algorithm 1} Step 4 to find \textbf{$(\hat{\mathbf{F}}_{P},\hat{\mathbf{W}}_{P},\hat{\mathbf{F}}_{B}[k],\hat{\mathbf{W}}_{B}[k])$}. 

In Fig. \ref{fig: orthogonal_ref_vs_eigen}, we can find that when
$\text{SNR}>0\text{ dB}$, the observations with and without noise
effect in the proposed method yield almost the same throughput. Then,
to better compare our approach with $I_{Ref}$, let us see curves
$I_{Pro}(Eig,N\!F,M=3)$ and \textit{$I_{Ref}$.} It is obvious that
\textit{$I_{Pro}(Eig,N\!F,M=3)$} achieves higher data rates than
\textit{$I_{Ref}$}. Although these two methods use different ways
to construct the hybrid beamforming, we try an explanation based on
some assumptions. Assume that these two schemes find the same $N_{RF}$
analog beam pairs, which means that they have the same effective channel.
In this case, \textbf{Algorithm 1} uses the SVD of the effective channel
to find the solution of digital beamforming matrices. From \cite{Telatar1999},
we know that this solution is the optimal. In contrast, the digital
beamforming in \cite{Ayach2012} uses the least-squares solution,
which is sub-optimal. When we reserve more candidates ($M>N_{RF}$),
there is a higher probability that both algorithms find the same $N_{RF}$
analog beam pairs. If so, \textbf{Algorithm 1} theoretically outperforms
the reference method. 

Next, in Fig. \ref{fig: orthogonal_eigen_vs_key}, when $\text{SNR}<5\text{ dB}$,
\textit{$I_{Pro}(Fro,M=3)$} achieves almost the same data rates as
\textit{$I_{Pro}(Eig,M=3)$}, which means that the approximation error
between $\sum_{n_{s}=1}^{N_{S}}\log_{2}(1+\gamma[\hat{\boldsymbol{\mathbf{\Sigma}}}_{E,i}^{2}[k]]_{n_{s},n_{s}})$
and $||\hat{\mathbf{H}}_{E,i}[k]||_{F}^{2}$ (see (\ref{eq: selection_criterion}))
is small in the low SNR regime. From Fig. \ref{fig: orthogonal_ref_vs_eigen}
and Fig. \ref{fig: orthogonal_eigen_vs_key}, we can see that if the
system operates in the SNR range between $-5$ and $5$ dB, the Frobenius
norm of the estimated effective channel works pretty well.

\section{Conclusion}

This paper presents a novel strategy to implement the hybrid beamforming
matrices at the transmitter and receive based on the received coupling
coefficients. As a result, the high-dimensional channel estimation
and singular value decomposition are unnecessary. The idea behind
this approach is simple: efficiently evaluating the key parameter,
such as the Frobenius norm of the effective channel matrices, to implement
the hybrid beamforming based on the estimates of received power levels.
Since the key parameter of the hybrid beamforming gain is a function
of the effective channel matrix, which has much lower dimension typically,
it is not difficult to try a (small) set of possible alternatives
to find a reasonable approximation of the optimal hybrid beamforming.
Moreover, the effective channel matrix can be obtained from the received
coupling coefficients. This avoids acquiring the \textit{explicit}
channel estimation and knowledge of the specific angles of the propagation
paths. Instead, the \textit{implicit} channel knowledge that which
beam pairs produce the strongest coupling between transmitter and
receiver is enough in a sense. 

\section*{Acknowledgment}

The research leading to these results has received funding from the
European Union\textquoteright s Horizon 2020 research and innovation
programme under grant agreement No. 671551 (5G-XHaul) and the TUD-NEC
project \textquotedblleft mmWave Antenna Array Concept Study\textquotedblright ,
a cooperation project between Technische Universität Dresden (TUD),
Germany, and NEC, Japan.

\bibliographystyle{IEEEtran}
\bibliography{reference}

\end{document}